\documentclass[sigconf=]{acmart}
\AtBeginDocument{%
  }

\setcopyright{acmlicensed}
\copyrightyear{2026}
\acmYear{2026}
\acmDOI{XXXXXXX.XXXXXXX}
\acmISBN{978-1-4503-XXXX-X/2018/06}

\usepackage{amsmath} 
\usepackage{amsfonts} 
\usepackage{mathrsfs}
\usepackage{enumitem}
\usepackage{booktabs}  
\usepackage{tcolorbox}
\usepackage{multirow} 
\usepackage{graphicx}  
\usepackage{amsmath}  
\usepackage{mathrsfs}  
\usepackage{subcaption}
\usepackage{caption}
\usepackage{subcaption}

\usepackage{algorithm}
\usepackage{algorithmic}
\allowdisplaybreaks

\begin{document}

% \title{Collaboration and Confrontation: Infusing Personalized Knowledge from Large Language Models into Recommendation}
\title{LWGR: Lagrangian-Constrained Personalized World Knowledge for Generative Recommendation}

\author{Lingyu Mu}
% \orcid{0000-0002-7252-5207}
\authornotemark[1]
\affiliation{
  \institution{Institute of Information Engineering, Chinese Academy of Sciences}
  \city{Beijing} 
  \state{} 
  \country{China}
}
\email{mulingyu@iie.ac.cn}

\author{Hao Deng}
\orcid{0009-0002-6335-7405}
\authornote{Contributed equally to this research.} 
\affiliation{%
  \institution{Alibaba International Digital Commerce Group}
   \city{Beijing} 
   \state{} 
   \country{China}
}
\email{denghao.deng@alibaba-inc.com}

\author{Haibo Xing}
\orcid{0009-0006-5786-7627}
\affiliation{%
  \institution{Alibaba International Digital Commerce Group}
  \city{Hangzhou} 
  \state{} 
  \country{China}
}
\email{xinghaibo.xhb@alibaba-inc.com}

\author{Kaican Lin}
\affiliation{%
  \institution{Alibaba International Digital Commerce Group}
  \city{Beijing} 
  \state{} 
  \country{China}
}
\email{linkaican.lkc@alibaba-inc.com}

\author{Zhitong Zhu}
\affiliation{
  \institution{Institute of Information Engineering, Chinese Academy of Sciences}
  \city{Beijing} 
  \state{} 
  \country{China}
}
\email{zhuzhitong@iie.ac.cn}

\author{Yu Zhang}
\orcid{0000-0002-6057-7886}
\affiliation{
  \institution{Alibaba International Digital Commerce Group}
  \city{Beijing} 
  \state{} 
  \country{China}
}
\email{daoji@lazada.com}

\author{Xiaoyi Zeng}
\orcid{0000-0002-3742-4910}
\affiliation{
  \institution{Alibaba International Digital Commerce Group}
  \city{Hangzhou} 
  \state{} 
  \country{China}
}
\email{yuanhan@taobao.com}

\author{Zhengxiao Liu$^\dag$}
\affiliation{
  \institution{Institute of Information Engineering, Chinese Academy of Sciences}
  \city{Beijing} 
  \state{} 
  \country{China}
}
\email{liuzhengxiao@iie.ac.cn}

\author{Zheng Lin}
\authornote{Corresponding authors.}
\affiliation{
  \institution{Institute of Information Engineering, Chinese Academy of Sciences}
  \city{Beijing} 
  \state{} 
  \country{China}
}
\email{linzheng@iie.ac.cn}

\author{Jinxin Hu}
\orcid{0000-0002-7252-5207}
\affiliation{
  \institution{Alibaba International Digital Commerce Group}
  \city{Beijing} 
  \state{} 
  \country{China}
}
\email{jinxin.hjx@alibaba-inc.com}

\begin{abstract}
Recent progress in large language model (LLM) based generative recommendation (GR) shows that leveraging LLM world knowledge can substantially improve performance. However, existing methods rely on fixed, manually designed instructions to generate semantic knowledge and directly incorporate it into GR, which has two limitations: (1) fixed instructions cannot capture the multidimensional heterogeneity of user interests; (2) uncontrollable knowledge fusion may conflict with behavioral signals and harm recommendations.
To address these limitations, we propose \textbf{LWGR}, a framework that leverages \textbf{L}agrangian constraints to transfer users' personalized \textbf{W}orld knowledge from LLMs into \textbf{G}enerative \textbf{R}ecommendation. LWGR enhances GR along two axes: knowledge extraction and fusion. It builds user personalized soft instructions to extract behavior-relevant LLM world knowledge. Then, it formulates knowledge fusion as an optimization problem with explicitly bounded performance degradation, solved via a Lagrangian primal–dual method that selectively incorporates beneficial knowledge.
We further design two training strategies for different LLM scales and a deployment scheme that combines nearline precomputation with lightweight online serving. 
Experiments on multiple public datasets and one industrial dataset show that LWGR outperforms eight state-of-the-art baselines by up to 11.23\% and brings a 1.35\% revenue lift on a large-scale advertising platform, demonstrating its effectiveness and practicality.

\end{abstract}


\vspace{-3pt}
\vspace{-3pt}


\maketitle
\begin{figure}[t]
  \centering
  \includegraphics[width=1\linewidth]{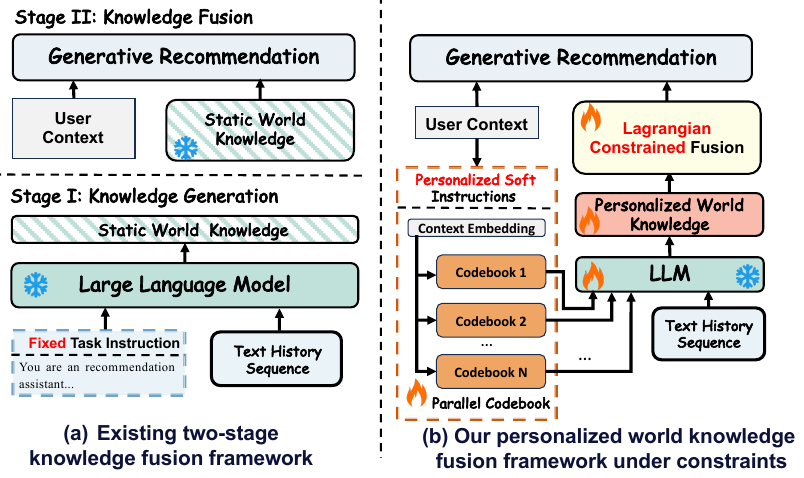}
  \caption{Existing two-stage knowledge fusion GR based on fixed task instruction and our GR that fuses user-personalized world knowledge under constraints. }
  \label{fig1}
  \vspace{-0.55cm} %调整图片与上文的垂直距离
\end{figure}
\section{Introduction}
Recently, generative recommendation (GR) \cite{tiger, cobra} has reduced the reliance on large-scale item embedding tables \cite{wang2021survey, wang2024rethinking, lin2024enhancing,wang2025home, mu2025trust} and demonstrated competitive performance by directly generating a set of discrete tokens of target items (i.e., SIDs). However, its semantic modeling ability is mainly limited to users’ historical interactions, making it difficult to explicitly leverage richer external knowledge for semantic association mining, and thus it struggles to capture complex interest structures.
% In recent years, rapidly developing large language models (LLMs) have effectively alleviated these issues by leveraging world knowledge.
% With the exponential growth of model scale and training corpora, LLMs have demonstrated the ability to implicitly encode world knowledge through self-supervised learning \cite{achiam2023gpt,bai2023qwen,liu2024deepseek,touvron2023llama}.
With the exponential growth of model scale and training corpora, large language models (LLMs) can extract world knowledge from massive text data, including factual information, commonsense reasoning, and abstract concepts, and store it in distributed parameters \cite{yu2023kola,guu2020retrieval}. This world knowledge significantly enhances their understanding and generalization capabilities \cite{achiam2023gpt,bai2023qwen,liu2024deepseek,touvron2023llama,yang2024unifying}.
Therefore, how to efficiently and controllably incorporate the world knowledge embedded in LLMs into GR, so as to systematically enhance recommendation capability is gradually becoming an important research direction \cite{zhao2024recommender, liu2023chatgpt, wang2024rethinking}.
% Therefore, how to efficiently and controllably incorporate the world knowledge embedded in LLMs into the GR has become an important research direction\cite{zhao2024recommender, liu2023chatgpt, wang2024rethinking}. The goal is to systematically enhance recommendation capability rather than relying solely on behavioral signals.

Existing works typically adopt a prompt-based two-stage knowledge fusion paradigm \cite{ren2024enhancing, KAR, lin2024clickprompt}. A prompt usually consists of two parts: a \textbf{task instruction} that guides the LLM to perform recommendation-related reasoning, and a \textbf{task data} instantiated with user profiles, historical behavior sequences, and multimodal item features. Together, they form a structured prompt template fed into the LLM to generate semantically enriched implicit knowledge representations, which are then integrated into GR via deep neural networks or attention mechanisms.
% Most existing studies focus on how to exploit such semantic knowledge, but lack a systematic analysis of how it is generated. 
Most existing studies \cite{KAR, ren2024enhancing} focus on how to exploit such semantic knowledge, but pay less attention to how to generate user-aligned knowledge in the first place. 
Recent work \cite{xu2025tapping,zhang2024causality} shows that the instruction part largely determines which world knowledge is invoked and how useful the output is and discrete identifiers such as user IDs and item IDs cannot be directly understood by LLMs.
Motivated by this, we conduct pilot experiments in real-world e-commerce scenarios to examine how different instructions affect the elicited semantic knowledge, and obtain two key findings: \textbf{(1)} The consistency between task instructions and user characteristics critically influences the value of world knowledge. Customized instructions for specific user groups can substantially raise the upper bound of GR performance. 
\textbf{(2)} 
% Misaligned instructions can induce knowledge that degrades recommendation performance.
Not all world knowledge is beneficial to GR. When instructions are misaligned with users’ interest structures, the induced knowledge can degrade performance.

Building on the above findings, we identify two limitations of the existing paradigm. \textbf{First}, a unified designed instruction cannot adapt to diverse user interest structures and thus underutilizes the world knowledge in LLMs. \textbf{Second}, the process of integrating knowledge into GR lacks explicit constraints, so inconsistencies between knowledge and behavioral signals can cause uncontrollable performance degradation. These issues raise two research questions:
\textbf{(i) How can we enable LLMs to generate personalized world knowledge that is highly consistent with users’ behavioral patterns?}
\textbf{(ii) How can we constrain the fusion process of world knowledge so that it can stably improve GR?}

To address the above issues, we propose LWGR, a framework that leverages \textbf{L}agrangian constraints to transfer users’ personalized \textbf{W}orld knowledge from LLMs into \textbf{G}enerative \textbf{R}ecommendations (Figure~\ref{fig1}(b)). LWGR improves the performance upper bound of GR from both the knowledge extraction and fusion perspectives:
\begin{itemize}[noitemsep, topsep=0pt, leftmargin=*]
        \item \textbf{Knowledge Extraction. }To construct personalized world knowledge, a naive solution is to manually design user personalized semantic instructions, but this is infeasible to build and maintain at scale. Inspired by soft prompts \cite{qin2021learning}, we instead learn soft instructions tightly coupled with user context, using continuous vectors to encode personalized preferences. 
        % Building on codebook-based item quantization, which represents multi-dimensional item characteristics via combinations of codewords, we transfer this idea to the user side by applying Optimized Product Quantization (OPQ) \cite{ge2013optimized} to user context vectors and clustering their interest structures in multiple subspaces. 
        Instead of directly assigning learnable instruction vectors to each user, this approach suffers from poor scalability and can only represent user interests in a linear space. Inspired by the codebook concept, we apply optimized product quantization (OPQ) \cite{ge2013optimized} to the user context vector, dividing it into multiple subspaces, and using combinations of selected codewords from a parallel codebook as personalized soft instructions.
        The selected codewords are combined into personalized soft instructions, concatenated with the task data, and fed into the LLM to obtain user personalized world knowledge. To enable end-to-end training over the entire codebook space, we further adopt a straight-through gradient mechanism based on Index Backpropagation Quantization (IBQ) \cite{shi2025scalable}. 
    \item \textbf{Knowledge Fusion. }After obtaining user-personalized world knowledge, we integrate it at the beginning of GR decoding via cross-attention to globally guide autoregressive generation. To ensure stable gains, we further formulate knowledge fusion as an optimization problem with an explicit upper bound on performance degradation, and use a Lagrangian primal–dual method to dynamically penalize performance loss, guiding the model to selectively integrate beneficial knowledge.
\end{itemize}
Meanwhile, we design two training strategies for LLMs of different scales, and propose a deployment scheme that combines nearline precomputation with lightweight online querying.

% We conduct extensive assessments of LWGR across two public datasets and one industry-scale real-world dataset, comparing its performance against eight state-of-the-art (SOTA) baselines (\textit{e.g.}, KAR \cite{KAR}, SeRALM \cite{ren2024enhancing}, TIGER \cite{tiger}). LWGR consistently achieves the best performance in all comparison scenarios, demonstrating up to 11.23\% improvement over the base models in recommendation performance. Furthermore, we have deployed LWGR on a large-scale commercial advertising platform, where online A/B testing revealed a 1.35\% increase in advertising revenue and 1.17\% growth in click-through rate (CTR). These empirical results demonstrate the significant practical implications of LWGR for real-world RSs.
We evaluate LWGR on two public benchmarks and one large-scale industrial dataset, and compare it with eight state-of-the-art (SOTA) baselines, including KAR \cite{KAR}, SeRALM \cite{ren2024enhancing}, and TIGER \cite{tiger}. LWGR consistently delivers the best results across all settings, with the largest relative improvement reaching 11.23\% over the strongest baseline. We further deploy LWGR on a large-scale commercial advertising platform, where an online A/B test shows a 1.35\% lift in advertising revenue and a 1.17\% increase in click-through rate (CTR). These offline and online results highlight the practical value of LWGR for real-world recommender systems.

Our contributions can be summarized as follows:
\begin{itemize}[noitemsep, topsep=0pt, leftmargin=*]
    \item We propose LWGR, which transfers users’ personalized world knowledge in LLMs into GR under Lagrangian constraints.
    \item We design a personalized knowledge extraction module based on parallel codebooks to integrate user personalized world knowledge into decoding, and model this fusion as a bounded-degradation optimization problem solved via a Lagrangian primal–dual method.
    \item We enable end-to-end training with an IBQ-based straight-through gradient mechanism and propose two training strategies plus a lightweight deployment scheme for LLMs of different scales.
    \item We achieve SOTA results on multiple public benchmarks and a real-world industrial dataset, with up to 11.23\% improvement over eight SOTA recommendation baselines, demonstrating the effectiveness of LWGR.
\end{itemize}

\section{Related Work}
\subsection{Generative Recommendation}
Unlike traditional discriminative recommenders \cite{morec,embedding-1,embedding-2,embedding-3,embedding-4}, GR treats recommendation as a generative modeling problem \cite{wang2023generative,rajput2023recommender}. GR first encodes item content (\textit{e.g.}, text, images) into a continuous semantic representation, and then performs vector quantization to obtain discrete SIDs \cite{hua2023index,hou2025generative,wu2024survey}. TIGER \cite{tiger} is an early SID‑based GR model that employs an RQ‑VAE \cite{lee2022autoregressive} to quantize text features and uses an autoregressive Transformer \cite{vaswani2017attention} to generate SIDs token by token. Cobra \cite{cobra} enriches SIDs with continuous representations and alternates between predicting discrete codes and continuous embeddings, thereby narrowing the gap between generation and retrieval. RPG \cite{rpg} adopts optimized product quantization \cite{ge2013optimized} to enhance codebook capacity by partitioning an item’s semantic vector into multiple subspaces and quantizing each separately, yielding longer SIDs with more fine‑grained semantics. Reg4Rec \cite{reg4rec} further introduces reinforcement learning to strengthen the reflective and self‑correction abilities of GR, enlarging the exploration space of generation trajectories and lifting its attainable performance.

\subsection{Prompt-Based Two-Stage Fusion in GR}
To effectively utilize the reasoning capabilities of LLMs grounded in world knowledge \cite{hagoort2004integration,csmf, arkin1990integrating}, recent works have introduced a prompt-based two-stage knowledge fusion paradigm. 
For example, KAR \cite{KAR} constructs user-side preference prompts and item-side factual prompts, feeding them into LLMs to obtain personalized reasoning and factual knowledge. These two types of knowledge are then passed through a knowledge adaptation module based on a mixture-of-experts (MoE) \cite{jacobs1991adaptive} framework, which compresses the outputs into low-dimensional representations compatible with recommendation features before integration into the recommendation model.  
Similarly, SeRALM \cite{ren2024enhancing} first designs prompts to guide the LLM to generate text-based knowledge, which is then encoded into vectors by a text encoder. In the second stage, the recommendation model is trained using both the semantic vectors from the first stage and the original ID embeddings.

\section{Preliminaries}
\subsection{Generative Recommendation}
\label{sec:GR}
Let $\mathcal{I}$ and $\mathcal{U}$ denote the item and user sets, respectively. Each user $u \in \mathcal{U}$ is associated with a historical interaction sequence $s_u = (i_1, i_2, \ldots, i_T)$, where $i_{t}$ is the item interacted with at step $t$ and $T$ is the maximum sequence length. In GR, the retrieval task is cast as a sequence generation task: conditioned on \(s_u\), the model produces a discrete identifier that corresponds to the target item. Concretely, we consider $L$ discrete vocabularies $\{V_\ell\}_{\ell=1}^L$. Each vocabulary $V_\ell$ is linked to a codebook matrix $C_\ell \in \mathbb{R}^{|V_\ell|\times d_\ell}$, where each row is a codeword indexed by an element in $V_\ell$. With these codebooks, every item $i$ is encoded as a length-$L$ token sequence $c_i = (c_i^1, \ldots, c_i^L)$, where $c_i^\ell \in V_\ell$ specifies the index of the selected codeword in the $\ell$-th codebook $C_\ell$. The learning goal of GR is to model the conditional distribution $p_{\theta}(c \mid s_u)$, i.e., to generate a SID $c$ given the user history $s_u$, and then retrieve the corresponding concrete item $i$. During training, for a user $u$ and the next-item label $i^+$, we take $c_{i^+}$ as the supervision signal and maximize the conditional log-likelihood $\max_{\theta}\mathbb{E}_{(u,i^+)}\big[\log p_{\theta}(c_{i^+}\mid s_u)\big].$
% \[
% \max_{\theta}\mathbb{E}_{(u,i^+)}\big[\log p_{\theta}(c_{i^+}\mid s_u)\big].
% \]
Under an autoregressive factorization, this conditional distribution can be written as
\begin{small}
\begin{equation}
p_{\theta}(c_{i^+}\mid s_u)= \prod_{\ell=1}^{L} p_{\theta}\big(c_{i^+}^{\ell} \mid c_{i^+}^{<\ell}, s_u\big),
\label{eq:factorize}
\end{equation}
\end{small}
and training reduces to next-token prediction by minimizing the negative log-likelihood of the ground-truth SID tokens:
\begin{small}
\begin{equation}
\mathcal{L}_{rec}= -\sum_{\ell=1}^{L} \log p_{\theta}\big(c_{i^+}^{\ell} \mid c_{i^+}^{<\ell}, s_u\big).
\label{eq:main_loss}
\end{equation}
\end{small}

\begin{figure}[t]
  \includegraphics[width=0.49\textwidth]{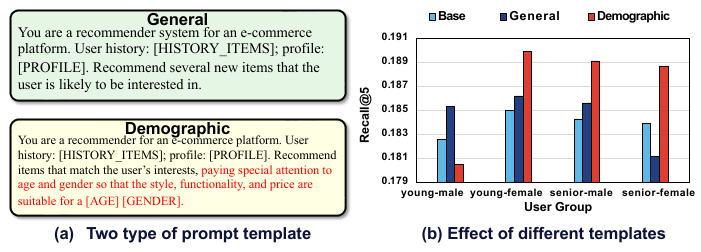}
  \caption{(a) General and demographic-guided instruction templates. (b) Performance of Base, General, and Demographic instructions across 4 user groups.}
  \label{fig:prompt_data}
  \vspace{-0.5cm}
\end{figure}

\subsection{Pilot Experiments}
Existing studies mostly focus on how to use LLM-generated semantic knowledge, rather than how to generate higher-quality knowledge in the first place. Recent work shows that the task instruction in the prompt strongly affects which world knowledge is activated. Motivated by this, we conduct a pilot study in a real-world e-commerce scenario to examine how different instructions shape the usefulness of world knowledge for recommendation. We split users into four cohorts by age and gender (young-male, young-female, senior-male, senior-female) and randomly sample 10k users from each. We use TIGER \cite{tiger}, a GR model without world knowledge, as the base model. Following prior work \cite{KAR}, we query an LLM with prompts constructed from users’ clicked item titles, take the pooled embedding of its outputs, and inject this vector into the encoder as the representation of the first token. As shown in Figure \ref{fig:prompt_data}(a), we design two prompt templates with the same task data but different instructions: (i) a general instruction that only specifies the recommender role and task, and (ii) a demographic-guided instruction that additionally asks the LLM to pay special attention to age and gender when generating recommendation-related knowledge. 
% Figure~\ref{fig:prompt_data}(b) reports the results for four user groups (Group~1–4 = young-male, young-female, senior-male, senior-female). 
For each group, we evaluate three variants (Base, General, Demographic), run each setting 5 times, and report the average Recall@5 \cite{xing2025esans}. From Figure \ref{fig:prompt_data}(b), we observe:
\begin{itemize}[noitemsep, topsep=0pt, leftmargin=*]
\item \textbf{Demographic-guided instructions can raise the performance ceiling.} Compared with the general instruction, the demographic-guided prompt achieves larger gains in several groups, indicating that when the instruction matches the user group, the semantic knowledge produced by the LLM is more beneficial.
\item \textbf{Not all world knowledge is beneficial.} In the young-male group, the demographic-guided instruction performs worse than the general instruction and even falls below the base model, suggesting that the demographic priors encoded in the LLM are misaligned with real user behavior and may interfere with behavioral signals when injected directly.
\end{itemize}
These findings suggest that an ideal solution should customize instructions to user context so that the LLM selectively activates user‑relevant world knowledge. However, manually crafting per‑user natural‑language instructions is infeasible at scale. Inspired by soft prompts, we instead adopt soft instructions and learn user‑specific instruction representations end‑to‑end. Moreover, knowledge integration must be controllable so that the model can suppress harmful knowledge and ensure stable gains.

\begin{figure*}[ht]
  \centering
  \includegraphics[width=1.0\textwidth]{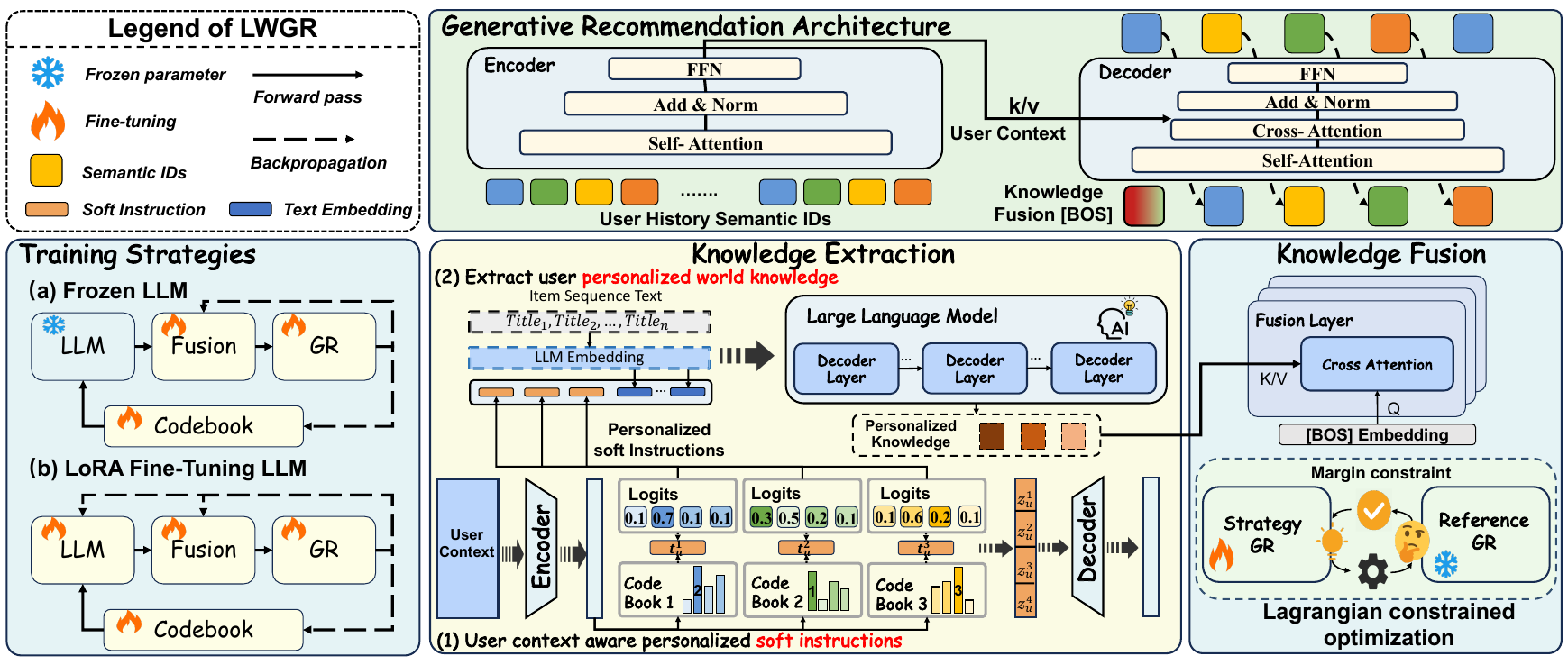 } % Reduce the figure size so that it is slightly narrower than the column.
  \caption{Overview of LWGR: a framework that integrates user personalized world knowledge into GR under Lagrangian constraints. LWGR first applies OPQ to quantize user context and uses parallel codebooks to generate personalized soft instructions. These instructions are concatenated with textualized historical behaviors and fed into the LLM to obtain user personalized world knowledge. This knowledge is injected into the \texttt{[BOS]} vector via cross-attention so that it influences the entire decoding process. The fusion is modeled as a constrained optimization problem to adaptively select beneficial knowledge. LWGR supports two training strategies: frozen parameters and LoRA-based fine-tuning, for LLMs of different scales and application scenarios.
  }
    \label{fig:3}
\end{figure*}
\section{Method}
\subsection{Overview}
This section introduces the proposed LWGR framework shown in Figure \ref{fig:3}, from \textbf{four aspects}.
(1) \textbf{Knowledge extraction}. Inspired by parallel codebooks and soft prompts, we select the most relevant codewords from multiple sub-codebooks for each user to form a non-linguistic personalized soft-instruction sequence. This code is concatenated with item text and fed into the LLM to extract personalized world knowledge related to the current user context.
(2) \textbf{Knowledge fusion}. We formulate the knowledge fusion process as a constrained optimization problem and use a Lagrangian primal–dual method to adaptively adjust the penalty on performance degradation. This highlights and amplifies knowledge that truly benefits the recommendation task while preserving the original recommendation ability.
(3) \textbf{Training strategies}. To seamlessly integrate the world knowledge fusion module into GR, we provide two training strategies: a freezing strategy that only trains the personalized knowledge module, and a lightweight LoRA-based fine-tuning strategy \cite{hu2022lora}. The model can flexibly choose according to LLM scale and computational budget.
(4) \textbf{Online deployment}. Finally, we plug LWGR into existing GR frameworks with minimal intrusion, combining offline precomputation with lightweight online serving to deliver real-time knowledge-enhanced recommendation for each user under practical latency and resource constraints.
% At last, we minimally intrusively plug LWGR into existing GR frameworks, combining offline precomputation with lightweight online calls to deliver real-time knowledge-enhanced recommendation for each user under practical latency and resource constraints.

% \subsection{End-to-end personalized knowledge extraction with parallel codebooks}
\subsection{Knowledge Extraction}
Our pilot experiments show that instructions with different orientations lead to markedly different gains across user groups, indicating that fixed hand-crafted templates are difficult to adapt to heterogeneous user preferences. Therefore, we adopt an end-to-end learning framework so that the model can automatically derive a soft instruction from a shared semantic space for each user, matching their behavioral patterns and guiding the LLM to extract the most relevant world knowledge.
\subsubsection{\textbf{User context aware personalized soft instructions.}} Based on the pilot experiments, a straightforward solution is to assign $K$ learnable vectors to each user as soft instructions. However, this makes the parameter size grow linearly with the number of users and prevents structure sharing across users. It also restricts each user to express only a limited set of interest patterns with a linear number of vectors. In contrast, parallel codebooks select codewords from $K$ subspaces of size $m$ and combine them into instructions, expanding the representational space from linear to $K^m$ under a similar parameter budget \cite{van2017neural,kingma2013auto,lee2022autoregressive,dong2020hawq}. This enables cross-user sharing and much richer interest modeling, making it better suited for large-scale personalization. Thus, we project user context vectors into multiple subspaces and perform codebook-based quantization within each subspace to obtain a set of discrete embeddings. These embeddings are treated as personalized soft instructions that best fit the user’s behavioral patterns.

Concretely, given the user context representation of user $u$ produced by the encoder of GR, denoted as $\mathbf{H}_{end} \in \mathbb{R}^{T \times d}$, we first apply mean pooling over the temporal dimension to obtain a user vector $\mathbf{h}^k \in \mathbb{R}^d$, and then project it into $K$ subspaces:
\begin{equation}
\mathbf{u}^k = f_k(\mathbf{h}^k) \in \mathbb{R}^{d_k}, \quad k = 1, \ldots, K,
\end{equation}
where $f_k(\cdot)$ is the linear projection layer for the $k$-th subspace and $d_k$ is its dimensionality. For each subspace $k$, we maintain a codebook matrix $\mathbf{C}_k \in \mathbb{R}^{|V_k| \times d_k}$, where $|V_k|$ is the index set of codewords and each row $\mathbf{C}_k[j]$ is a candidate codeword. We then quantize $\mathbf{u}^k$ in each subspace to its nearest codeword:
\begin{equation}
c_u^k = \arg\min_{j \in V_k} \left\| \mathbf{u}^k - \mathbf{C}_k[j] \right\|_2^2,
\end{equation}
where $c_u^k$ denotes the index selected for user $u$ in the $k$-th subspace. Collecting the indices from all subspaces yields the personalized discrete tokens of user $u$, i.e., $c_u = (c_u^1, \ldots, c_u^K)$. After obtaining $c_u$, a natural idea is to follow the practice on item codebooks: treat the discrete encoding $c_u$ as a set of special tokens, concatenate them with the item text tokens, and feed the whole sequence into the LLM, so that the model can explicitly perceive the user context when generating world knowledge. However, directly feeding discrete index $c_u$ or their codeword vectors $C_k[j]$ into the LLM without a differentiable selection mechanism breaks end-to-end learning. The $\arg\min$ operation used to select codewords is non-differentiable, so during backpropagation the recommendation loss can only update the chosen codeword and cannot shape the whole codebook $C_k$. This conflicts with our goal, since we want the loss to jointly learn all candidate codewords so that the encoding space is organized in a self-adaptive, end-to-end manner.

% In this way, we construct a candidate set of size $K$, where the encodings in each subspace span a relatively independent semantic dimension, and the combination of encodings across subspaces determines the multi-view hints for this user, providing the basis for subsequent personalized world knowledge extraction.
% We derive personalized soft instructionss by selecting one codeword from each of $K$ parallel sub-codebooks, therefore, $K$ is also the number of personalized soft instructions tokens prepended to the LLM input. A direct hard assignment via $\arg\max$ is non-differentiable: if we index the codebook using a hard index, the next-item cross-entropy loss only updates the selected codeword, while the code selection mechanism receives no gradient signal. This makes codebook learning overly sensitive to initialization and slows down the adaptation of the personalized prompts.

To enable end-to-end optimization under the GR objective, we adopt an IBQ-based \cite{shi2025scalable} straight-through estimator over the categorical distribution of codewords. Let the subspace projections $\{\mathbf{u}^k\}_{k=1}^K$ and the
codebooks $\{\mathbf{C}_k\}_{k=1}^K$ be defined as in Eqs.~(3)--(4).
We then compute similarity scores between $\mathbf{u}^k$ and all codewords and apply the straight-through estimator as follows:
\begin{equation}
\alpha_k^j=\mathrm{sim}(\mathbf{u}^k,\mathbf{C}_k[j]),
\qquad
p_j^k=\frac{\exp(\alpha_k^j/\tau)}{\sum_{j\in V_k}\exp(\alpha_k^j/\tau)},
\end{equation}
where $\tau>0$ is a softmax temperature and $\mathbf{p}^k=(p_k^j)_{j\in V_k}$, and $\mathrm{sim}(\cdot,\cdot)$ is the negative squared Euclidean distance \cite{lee2022autoregressive}. We then take a hard one-hot index $\mathbf{e}_{\mathrm{hard}}^k=\mathrm{OneHot}(\arg\max_{j \in V_k} p_k^j)$ and construct a straight-through index:
\begin{equation}
\mathbf{e}^k=\mathbf{e}_{\mathrm{hard}}^k-\mathrm{sg}[\mathbf{p}^k]+\mathbf{p}^k,
\end{equation}
where $\mathrm{sg}[\cdot]$ denotes the stop-gradient operator. This construction preserves hard quantization in the forward (i.e., selecting a single codeword), while allowing gradients to flow through the soft distribution $\mathbf{p}^k$ in the backward, so that all codewords can be updated.

Finally, we obtain the quantized subspace vector and map it to the LLM hidden space:
\begin{equation}
\mathbf{z}_u^k=(\mathbf{e}^k)^\top \mathbf{C}^k\in\mathbb{R}^{d_k},\qquad
\mathbf{t}_u^k=\mathbf{W}_{L}^k\mathbf{z}_u^k\in\mathbb{R}^{d_{\mathrm{LLM}}},
\end{equation}
where $\mathbf{W}_{L}^k\in\mathbb{R}^{d_{\mathrm{LLM}}\times d_k}$ is a learnable projection matrix.
We treat $\{\mathbf{t}_u^1,\ldots,\mathbf{t}_u^K\}$ as embeddings of $K$ special instruction tokens and prepend them to the item text tokens (prefix prompting) as the LLM input.
The LWGR is trained with the standard next-token cross-entropy loss (Eq. (\ref{eq:main_loss})), enabling the GR objective to optimize all codewords in each $\mathbf{C}^k$ via $\mathbf{p}^k$, rather than only the selected one.

\subsubsection{\textbf{World Knowledge Extraction based on personalized soft instructions}}
After obtaining the user-aware personalized soft instructions embeddings \( \mathbf{t}_u \), we use it to extract world knowledge from the LLM that is most relevant to the current user’s interaction history. Specifically, we concatenate the textual feature vectors (titles and descriptions) \([h^1_{\text{text}}, \ldots, h^T_{\text{text}}]\) corresponding to the user's historical behavior sequence $s_u$ with the user's personalized soft instructions, and construct the input to the LLM decoding layer as:
\begin{equation}
\mathcal{X}_u = [\mathbf{t}_u^1,\ldots, \mathbf{t}_u^K, \quad h^1_{\text{text}}, \ldots, h^T_{\text{text}}],
\end{equation}
where $h_{\text{text}}$ is obtained by encoding the textual feature of items in $s_u$ using the same LLM, reusing its token embedding layer.
Then, we feed $\mathcal{X}_u$ into the LLM for forward inference, obtaining a mixed semantic representation that incorporates both the user's behavioral
context and world knowledge:
\begin{equation}
\mathbf{H}_u = \text{LLM}(\mathcal{X}_u).
\end{equation}
In this process, the context-aware personalized soft instructions vector
$\mathbf{t}_u$ explicitly informs the LLM of the current user's behavior
pattern, thereby guiding the LLM to preferentially activate world knowledge and semantic associations that are consistent with the user's preferences when processing the historical item texts.

\subsection{Knowledge Fusion}
Using the parallel codebook and the personalized soft instructions, we extract world knowledge representations from the LLM that are highly related to each user’s preferences. However, according to pilot experiments, directly integrating such knowledge as input into GR can easily conflict with the original behavioral signals and even damage the basic recommendation capability. Therefore, we first design a fusion module based on cross-attention, and then formulate the knowledge fusion process as a constrained optimization problem, enabling controllable incorporation of world knowledge under explicitly enforced ranking constraints.

\subsubsection{\textbf{Knowledge Fusion Module based on Cross-Attention. }}After obtaining the personalized world knowledge $\mathbf{H}_u$, the key question is how to effectively incorporate it into the decoding process. We hope to integrate this knowledge at a position that can exert a global influence on subsequent generation while minimally disturbing the original model architecture. To this end, we fuse world knowledge at the beginning of decoding. The decoder’s \texttt{[BOS]} token typically carries the global semantic state of the sequence, so by adjusting its representation, the integrated knowledge can be propagated through subsequent autoregressive layers, imposing a unified semantic prior on the entire recommendation sequence.

Based on the above considerations, we denote the initial representation of the \texttt{[BOS]} token as $\mathbf{q}_0 \in \mathbb{R}^d$, and fuse it with $\mathbf{H}_u$ through a cross-attention. Specifically, $\mathbf{q}_0$ serves as the query, and $\mathbf{H}_u$ serves as the key and value in multi-head cross attention:
\begin{equation}
\tilde{\mathbf{q}}_0 = \mathrm{CrossAttn}(\mathbf{q}_0, \mathbf{H}_u).
\end{equation}
where $\mathrm{CrossAttn}$ denotes the standard multi-head attention:
\begin{equation}
\mathrm{CrossAttn}(\mathbf{q}_0, \mathbf{H}_u)
= \sum_{h=1}^{H} \mathrm{softmax}\!\left( \frac{Q_h K_h^\top}{\sqrt{d_h}} \right) V_h W^O,
\label{CrossAttn_bos}
\end{equation}
where $H$ is the number of attention heads and $d_h$ is the dimension of each head. The projections are defined as $Q_h = \mathbf{q}_0 W_h^Q,K_h = \mathbf{H}_u W_h^K,V_h = \mathbf{H}_u W_h^V.$ The resulting $\tilde{\mathbf{q}}_0$ can be viewed as a \texttt{[BOS]} representation enhanced with the user’s world knowledge. We then feed $\tilde{\mathbf{q}}_0$ as the initial input to the decoder, while keeping the original decoding process unchanged. By introducing the cross-attention fusion at the \texttt{[BOS]}, the world knowledge can globally influence the subsequent generation, while preserving the original decoder architecture so that it can be adapted to most GRs \cite{tiger, reg4rec, zhou2025onerec}.

\subsubsection{\textbf{Lagrangian-constrained optimization. }} 

Although we fuse LLM-generated world knowledge into GR, pilot experiments show that such knowledge is not always beneficial. Aggressive fusion can introduce signals that conflict with behavior evidence, resulting in retrieval degradation. To make knowledge fusion controllable, we cast training as an inequality-constrained optimization problem. Specifically, we first pretrain a GR model without knowledge fusion on the same training data and with the same architecture, and frozen it as a reference model, denoted by $\theta_{\mathrm{ref}}$, and train a knowledge-conditioned GR model (the policy model) with parameters $\theta$. During training, we explicitly constrain the policy model to not underperform the reference model in its ability to generate the ground-truth next-item SID sequence.

Following the GR formulation in Section~\ref{sec:GR}, given a user behavior sequence $s_u$ and the ground-truth next item $i^+$ with SID sequence $c_{i^+}=(c_{i^+}^1,\ldots,c_{i^+}^{L})$, the policy model defines the autoregressive likelihood $p_{\theta}(c_{i^+}\mid s_u)$ (Eq.~(\ref{eq:factorize})). We score an item by the mean token log-probability:
\begin{equation}
s_{\theta}(u,i^+)=\frac{1}{L}\sum_{\ell=1}^{L}\log p_{\theta}\!\left(c_{i^+}^{\ell}\mid c_{i^+}^{<\ell}, s_u\right),
\label{eq:score}
\end{equation}
which measures the model's average confidence in generating $c_{i^+}$ conditioned on $s_u$. The frozen reference model $\theta_{\mathrm{ref}}$ computes $s_{\mathrm{ref}}(u,i^+)$ in the same way.

\paragraph{(1) Reference-based confidence degradation constraint.}
To prevent knowledge fusion from reducing the generation confidence on the ground-truth next item, we define a degradation penalty:
\begin{equation}
C_{\mathrm{margin}}(u,i^+;\theta)=\max\big(0,\ s_{\mathrm{ref}}(u,i^+)-s_{\theta}(u,i^+)-\delta\big),
\label{eq:cmargin_point}
\end{equation}
where $\delta\ge 0$ is a tolerance margin. We then aggregate it over the training data:
\begin{equation}
C(\theta)=\mathbb{E}_{(u,i^+)\sim\mathcal{D}}\big[C_{\mathrm{margin}}(u,i^+;\theta)\big],
\end{equation}
where $\mathcal{D}$ denotes the training data set. Finally, we impose the inequality constraint $C(\theta)\le \varepsilon$, where $\varepsilon$ controls the maximum average degradation (typically set to $0$ or a small positive value).

\paragraph{(2) Inequality-constrained objective.}
Let $\mathcal{L}_{\mathrm{rec}}(\theta)$ denote the GR training loss in Eq.~(\ref{eq:main_loss}). Combining the main objective with the reference-based degradation constraint, we obtain the following inequality-constrained optimization problem:
\begin{equation}
\min_{\theta}\ \mathcal{L}_{\mathrm{rec}}(\theta)\quad \text{s.t.}\quad C(\theta)\le \varepsilon .
\label{eq:constrained_obj}
\end{equation}
This means that we only accept model parameters $\theta$ that minimize the recommendation loss while keeping the average degradation below the threshold. A naive way to handle the constraint is to use a manually weighted hinge penalty:
\begin{equation}
\mathcal{L}_{\mathrm{total}}(\theta)=\mathcal{L}_{\mathrm{rec}}(\theta)+\beta\,\max\big(0,\ C(\theta)-\varepsilon\big),
\label{eq:manual_penalty}
\end{equation}
where $\beta$ is a fixed hyperparameter. However, $\beta$ is hard to tune: if it is too small, the constraint becomes ineffective; if too large, the model may overemphasize the constraint and sacrifice the main recommendation objective.
% , the constraint becomes ineffective, and if it is too large, the model may overemphasize satisfying the constraint and sacrifice the main recommendation objective.

To avoid manual weighting, we adopt a Lagrangian dual method, which replaces the fixed coefficient with an adaptive dual variable learned during training, allowing the constraint strength to be adjusted automatically.

\paragraph{(3) Lagrangian primal--dual optimization.}
We introduce a non-negative Lagrange multiplier $\lambda$ and form the Lagrangian:
\begin{equation}
\mathcal{L}_{\mathrm{total}}(\theta,\lambda)=\mathcal{L}_{\mathrm{rec}}(\theta)+\lambda\big(C(\theta)-\varepsilon\big),
\end{equation}
The constrained problem in Eq.~(\ref{eq:constrained_obj}) can be reformulated as the following saddle-point objective:
\begin{equation}
\min_{\theta}\ \max_{\lambda\ge 0}\ \mathcal{L}_{\mathrm{total}}(\theta,\lambda).
\end{equation}
In practice, we adopt a primal--dual gradient method and alternately update $\theta$ and $\lambda$. With $\lambda$ fixed, we update $\theta$ by gradient descent:
\begin{equation}
\theta \leftarrow \theta-\eta_{\theta}\Big(\nabla_{\theta}\mathcal{L}_{\mathrm{total}}(\theta)+\lambda\,\nabla_{\theta}C(\theta)\Big).
\end{equation}
With $\theta$ fixed, we update $\lambda$ by projected gradient ascent:
\begin{equation}
\lambda \leftarrow \max\big(0,\ \lambda+\eta_{\lambda}\big(C(\theta)-\varepsilon\big)\big),
\label{eq:lambda}
\end{equation}
where $\eta_{\theta}$ and $\eta_{\lambda}$ are the learning rates. Intuitively, when $C(\theta)>\varepsilon$, $\lambda$ increases to penalize constraint violations more strongly. Otherwise, $\lambda$ decreases and training focuses more on minimizing the main GR loss. Using Lagrangian primal--dual optimization, training no longer relies on a manually tuned fixed constraint weight. Instead, the constraint strength is adaptively determined by data via $\lambda$, which helps prevent the policy model from degrading relative to the reference model, while still allowing the main recommendation objective to improve upon the reference baseline.
% to further improve performance beyond the reference baseline.

\subsection{Training Strategy for LLM}
% \subsubsection{\textbf{LLM Update Strategy.} }
To accommodate different LLM sizes and compute budgets, we provide two update strategies:

\textbf{(1) Frozen LLM.} We treat the pretrained LLM as a fixed knowledge source and keep all its parameters frozen during training. We only update (i) the personalized soft instructions module, (ii) the knowledge fusion module, and (iii) the GR model conditioned on the personalized knowledge.

\textbf{(2) LoRA-based adaptation.} When additional compute is available, we perform parameter-efficient fine-tuning by inserting LoRA adapters \cite{hu2021lora} into the LLM while keeping the original weights $\phi$ frozen. Concretely, we apply LoRA to selected linear layers (e.g., attention projections and FFN layers). For a weight matrix $W_{llm} \in\mathbb{R}^{d_{\text{out}}\times d_{\text{in}}}$, we parameterize the adapted weight as
\begin{equation}
W'_{llm} = W_{llm} + \Delta W_{llm},\qquad \Delta W_{llm} = AB,
\end{equation}
where $A\in\mathbb{R}^{d_{\text{out}}\times r}$ and $B\in\mathbb{R}^{r\times d_{\text{in}}}$ are trainable low-rank matrices with rank $r\ll \min(d_{\text{out}}, d_{\text{in}})$, and $W_{llm}$ remains frozen. This adds minimal overhead and can help reduce the mismatch between general-purpose LLM knowledge and recommendation-specific signals. In our experiments, we compare these two strategies to quantify the effect of LoRA adaptation over the frozen-LLM setting.

\begin{figure}[t]
    \vspace{-5pt}
    \captionsetup{skip=2pt} 
  \includegraphics[width=0.49\textwidth]{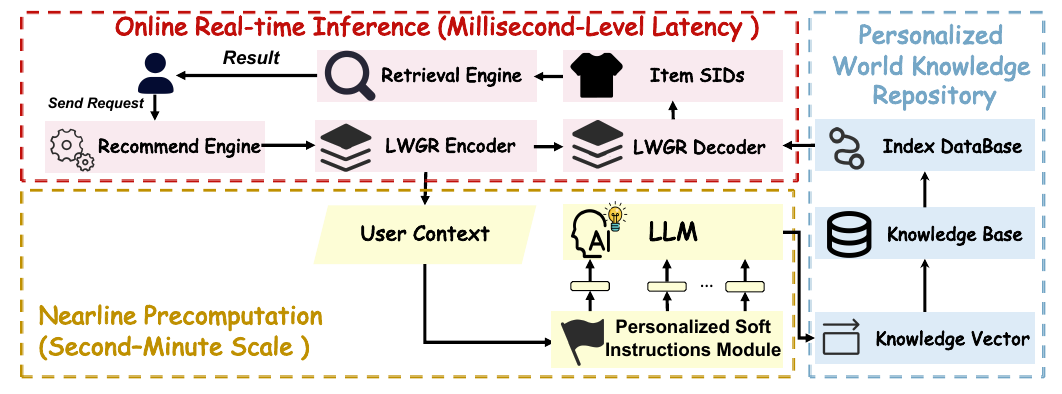}
  \caption{The online and nearline deployment of LWGR.}
  \label{fig:deployment}
    \vspace{-13pt}
\end{figure}

\subsection{Online Deployment}
\label{rec:online_deployment}
% To deploy the proposed LWGR in a production system, we adopt a novel hybrid strategy that combines nearline pre-computation with lightweight online inference. This design avoids online LLM execution and enables minimal changes to an existing GR pipeline.
% To deploy LWGR in a large-scale industrial recommender system, we adopt a hybrid serving pipeline that combines nearline precomputation with millisecond-level online inference (Figure~\ref{fig:deployment}). This design keeps all LLM serving in the nearline pipeline, where context–aware world knowledge is refreshed at a second–minute time scale, and only introduces a lightweight fusion step into the existing GR pipeline at request time. Compared with prompt-based two-stage knowledge fusion methods, this enables continuously updated personalized world knowledge while preserving millisecond-level online latency.
To deploy LWGR in a large-scale industrial recommender system, we adopt a hybrid serving pipeline that combines nearline precomputation with millisecond-level online inference (Figure~\ref{fig:deployment}). This design keeps all LLM serving in the nearline pipeline, where context-aware world knowledge is refreshed at a second–minute scale, and only adds a lightweight fusion step to the existing GR pipeline at request time. Compared with prompt-based two-stage knowledge fusion, this enables continuously updated personalized world knowledge while preserving millisecond-level latency.

\subsubsection{Nearline precomputation (second–minute scale)}
In the nearline pipeline, we continuously collect user context from the online LWGR encoder outputs. For each user $u$, we generate $K$ personalized soft instructions based on the latest context and feed them, together with the user’s item texts, into the LLM to extract a context-level world-knowledge representation $\mathbf{H}_u$. We then write $\mathbf{H}_u$ into a \textbf{personalized world knowledge repository}: an index database keyed by user ID that stores the latest knowledge vectors as user-level features. This repository is refreshed at a second–minute time scale (\textit{e.g.}, every few minutes) via incremental updates, keeping user knowledge up to date without any online LLM cost.

\subsubsection{Online real-time inference (millisecond-level latency)}
% When user $u$ issues a recommendation request, the online recommendation engine invokes the GR pipeline as usual. The LWGR encoder produces the current user context, and the service retrieves the latest cached $\mathbf{H}_u$ from the knowledge repository. This vector is then injected into the LWGR decoder through a lightweight cross-attention module, before the retrieval engine generates item candidates. Since the online pipeline only performs one index lookup and a single cross-attention block, the additional latency is on the order of milliseconds in practice. We further provide online evaluation results in Section~\ref{sec:online_Experiments}.
When user $u$ issues a request, the online recommendation engine invokes the GR pipeline as usual. The LWGR encoder produces the current user context, and the service retrieves the latest cached $\mathbf{H}_u$ from the knowledge repository. This vector is then injected into the LWGR decoder through a lightweight cross-attention module before the retrieval engine generates item candidates. Since the online pipeline only performs one index lookup and a single cross-attention block, the extra latency is only a few milliseconds in practice. Online evaluation results are reported in Section~\ref{sec:online_Experiments}.

Overall, this deployment scheme offers two practical advantages: (i) \textbf{effectiveness}: each request explicitly consumes the cached personalized world knowledge $\mathbf{H}_u$, which is continuously refreshed in the nearline pipeline at a second–minute time scale, enabling per-user, up-to-date knowledge-aware recommendations at serving time; and (ii) \textbf{system compatibility}: it reuses the existing retrieval engine and GR backbone, and only adds the context-aware knowledge vectors and a fusion layer, without requiring online LLM inference or intrusive architectural changes.

\section{Experiments}
To comprehensively evaluate LWGR’s performance, we conduct extensive experiments to address the following research questions:
\begin{itemize}[noitemsep, topsep=0pt, leftmargin=*]
    \item \textbf{RQ1}: Does LWGR outperform state-of-the-art retrieval models and two-stage knowledge fusion baselines?
    \item \textbf{RQ2}: Can scaling up LLM size enhance LWGR’s performance?
    \item \textbf{RQ3}: What are the contributions of each module in LWGR?
    \item \textbf{RQ4}: What impact do hyperparameters have on its effectiveness?
    \item \textbf{RQ5}: Can LWGR deliver performance improvements in real-world online recommendation platforms?
\end{itemize}

\subsection{Experimental Setup}
\subsubsection{Datasets and Evaluation Metrics.}
Our evaluation is conducted on three datasets: two derived from a public dataset and one large-scale industrial dataset. For the public data, we construct two subsets from the widely used Amazon Product Reviews collection \cite{AmazonDataset}. Specifically, we select the Beauty and Toys \& Games categories and, for each user, form an interaction sequence by chronologically ordering their reviews. Following common practice \cite{tiger}, we filter out users with fewer than five interactions to ensure data quality and meaningful sequence modeling. After filtering, Beauty contains about 22K users, 12K items, and 296K interactions, while Toys contains about 35K users, 18K items, and 167K interactions. The industrial dataset consists of internal interaction logs from a major Southeast Asian e‑commerce platform. It contains over 2.95B user–item interactions from 14.7M users and 20.1M items collected between January and July 2025, providing a realistic view of large‑scale user behavior in production. For each user, we record a behavior sequence including clicks and conversions, and each item is associated with rich multimodal content such as product images, titles, and textual descriptions. For evaluation, we employ a standard leave-one-out strategy \cite{sasrec}: for each user sequence, the most recent interaction is used for testing, the second most recent for validation, and the remaining interactions for training.

We evaluate recommendation performance using two standard metrics: Recall@5/10 (R@5/10) and NDCG@5/10 (N@5/10) \cite{tiger,unisrec}.

\begin{table*}[t]
\centering
\caption{Performance comparison on Beauty, Toys and Industrial datasets. Best results are in \textbf{bold} and second-best are \underline{underlined}. "\textbf{Improv.}" shows the relative improvement (\%) over the second-best method. }
\label{tab:beauty_toys_industry}
\resizebox{\textwidth}{!}{
\begin{tabular}{lcccccccccccc}
\toprule
\multirow{2}{*}{Methods} & \multicolumn{4}{c}{Beauty} & \multicolumn{4}{c}{Toys} & \multicolumn{4}{c}{Industry} \\
\cmidrule(lr){2-5}\cmidrule(lr){6-9}\cmidrule(lr){10-13}
& Recall@5  & NDCG@5 & Recall@10 & NDCG@10
& Recall@5  & NDCG@5 & Recall@10 & NDCG@10
& Recall@5  & NDCG@5 & Recall@10 & NDCG@10 \\
\midrule
\midrule
\multicolumn{13}{c}{\emph{Item ID-based Discriminative Recommendation}} \\
\midrule
SASRec        & 0.0512 & 0.0326 & 0.0861 & 0.0589 & 0.0543 & 0.0344 & 0.0764 & 0.0524 & 0.1637 & 0.1092 & 0.2092 & 0.1396 \\
PinnerFormer  & 0.0529 & 0.0335 & 0.0902 & 0.0619 & 0.0579 & 0.0371 & 0.0813 & 0.0568 & 0.1718 & 0.1145 & 0.2162 & 0.1447 \\
HeterRec      & 0.0521 & 0.0332 & 0.0894 & 0.0616 & 0.0574 & 0.0366 & 0.0811 & 0.0561 & 0.1763 & 0.1187 & 0.2219 & 0.1482 \\
VQ-Rec        & 0.0513 & 0.0322 & 0.0883 & 0.0608 & 0.0587 & 0.0372 & 0.0834 & 0.0573 & 0.1725 & 0.1154 & 0.2185 & 0.1464 \\
\midrule
\midrule
\multicolumn{13}{c}{\emph{Semantic ID-based Generative Recommendation}} \\
\midrule
TIGER         & 0.0534 & 0.0341 & 0.0912 & 0.0630 & 0.0612 & 0.0395 & 0.0865 & 0.0599 & 0.1842 & 0.1227 & 0.2317 & 0.1551 \\
Cobra         & 0.0527 & 0.0334 & 0.0909 & 0.0625 & 0.0634 & 0.0408 & \underline{0.0899} & \underline{0.0621} & 0.1821 & 0.1214 & 0.2345 & 0.1574 \\
TIGER + KAR   & \underline{0.0543} & \underline{0.0347} & 0.0927 & 0.0634 & 0.0629 & 0.0403 & 0.0886 & 0.0616 & 0.1893 & 0.1245 & 0.2422 & 0.1611 \\
TIGER + SeRALM& 0.0539 & 0.0344 & \underline{0.0931} & \underline{0.0642} & \underline{0.0637} & \underline{0.0411} & 0.0891 & 0.0619 & \underline{0.1945} & \underline{0.1273} & \underline{0.2467} & \underline{0.1639} \\
\midrule
\midrule
\multicolumn{13}{c}{\emph{Ours}} \\
\midrule
LWGR          & \textbf{0.0595} & \textbf{0.0376} & \textbf{0.1026} & \textbf{0.0701}
              & \textbf{0.0694} & \textbf{0.0443} & \textbf{0.0964} & \textbf{0.0669}
              & \textbf{0.2145} & \textbf{0.1416} & \textbf{0.2684} & \textbf{0.1804} \\
\textbf{Improv}. & +9.58\% & +8.36\% & +10.20\% & +9.19\% & +8.95\% & +7.79\% & +7.23\% & +7.73\% & +10.28\% & +11.23\% & +8.80\% & +10.07\% \\
\bottomrule
\end{tabular}}
\end{table*}

\subsubsection{Baseline Models.}
To comprehensively evaluate LWGR, we compare it against three groups of baselines: traditional ID-based recommenders, SID-based generative models, and prompt-based knowledge fusion methods. \textbf{ID-based recommenders.} \textbf{PinnerFormer} \cite{pancha2022pinnerformer} models long-range user behavior with a causally masked Transformer. \textbf{HeterRec} \cite{deng2025heterrec} adopts a dual-tower hierarchical Transformer for multi-modal items and a multi-step list-wise objective. \textbf{VQ-Rec} \cite{hou2023learning} builds an item codebook via VQ-VAE and represents items by aggregating selected codewords. \textbf{SASRec} \cite{sasrec} uses a unidirectional self-attention network for sequential recommendation. \textbf{SID-based GRs.} \textbf{TIGER} \cite{tiger} employs an RQ-VAE codebook and an encoder–decoder architecture to autoregressively generate semantic IDs. \textbf{Cobra} \cite{cobra} uses a decoder-only architecture that jointly models discrete SIDs and dense behavioral vectors. \textbf{Prompt-based knowledge fusion.} \textbf{KAR} \cite{KAR} elicits world knowledge via user-preference and item-factual prompts and then fuses the LLM outputs into the GR model through a MoE-style adaptation module. We instantiate KAR on top of TIGER, denoted as \textbf{TIGER+KAR}. \textbf{SeRALM} \cite{ren2024enhancing} uses fixed prompts to obtain text-based knowledge, encodes it into vectors, and trains the recommender jointly on these knowledge embeddings and item IDs; similarly, we integrate SeRALM with TIGER and report it as \textbf{TIGER+SeRALM}.
% our method with traditional ID-based recommendation models, generative recommendation methods, and the latest two-stage prompt-based knowledge fusion methods. \textbf{ID-based Recommendation:} \textbf{(1) Pinnerformer} \cite{pancha2022pinnerformer} relies on a causal-masked Transformer to characterize users’ long-range interaction histories. \textbf{(2) HeterRec} \cite{deng2025heterrec} leverages a dual-tower, hierarchical Transformer to encode multi-modal item information and is trained with a multi-step list-wise ranking objective. \textbf{(3) VQ-Rec} \cite{hou2023learning} constructs an item codebook using VQ-VAE and derives discrete item embeddings by aggregating the corresponding codewords.\textbf{(4) SASRec} \cite{sasrec} adopts a single-direction self-attention architecture to capture sequential user behavior. \textbf{Generative Recommendation:} \textbf{(1) TIGER} \cite{tiger}, which uses an RQ-VAE and an encoder-decoder architecture, and \textbf{(2) Cobra} \cite{cobra}, which employs a decoder-only architecture to fuse semantic IDs (from RQ-VAE) with dense behavioral vectors. \textbf{Prompt-based knowledge fusion methods:} \textbf{(1) KAR} \cite{KAR}, which elicits world knowledge via user‑preference and item‑factual prompts, and \textbf{(2) SeRALM} \cite{ren2024enhancing} first uses a fixed prompt template to elicit text-based world knowledge, then encodes the resulting texts into vectors and jointly trains the recommendation model on top of these knowledge embeddings together with item ID embeddings.

\subsubsection{Implementation Details.}
Our experiments are conducted on a distributed PyTorch \cite{paszke2019pytorch} platform with 2 parameter servers and 10 worker nodes, each equipped with one Nvidia A100 GPU. To ensure fairness, we use Qwen3-4B \cite{bai2023qwen} as the base LLM to extract semantic knowledge for all methods, including ours and the baselines. In our LWGR framework, the embedding dimension is set to 128 and the hidden size of Transformer layers is 640. For the parallel codebook used to generate personalized soft instructions, we set the number of codebooks to $K = 5$. For the LoRA fine-tuning strategy, we set the rank to 8, the scaling factor to 16, and the dropout rate to 0.05. LoRA updates are applied to the query and value projection matrices in the attention layers. We train all models with AdamW \cite{AdamW}, using a batch size of 32, an initial learning rate of $1\times 10^{-4}$. We apply gradient clipping with a maximum norm of 1.0 and use mixed-precision training with \texttt{bf16} \cite{bf16}  to accelerate training. For the inequality-constrained knowledge fusion objective, we initialize the Lagrange multiplier as $\lambda_0 = 0.05$. The model parameters are updated with learning rate $\eta_\theta = 1\times 10^{-4}$, while $\lambda$ is updated with a separate learning rate $\eta_\lambda = 5\times 10^{-4}$. We update $\lambda$ at every training step according to Eq.~(\ref{eq:lambda}). In the per-sample degradation penalty of Eq.~(\ref{eq:cmargin_point}), we set the margin $\delta = 1\times 10^{-4}$, so that any confidence drop relative to the reference model is counted as degradation. The constraint threshold $\varepsilon$ in Eq.~(\ref{eq:constrained_obj}) controls the maximal tolerable average degradation over the training data. In practice, we found that theLWGR is robust when $\delta$ and $\varepsilon$ are chosen within $[10^{-4}, 10^{-3}]$, and we report results with $\delta = \varepsilon = 1\times 10^{-4}$.
% In the per-sample degradation penalty of Eq.~\ref{eq:cmargin_point}, we set the margin $\delta = 1\times 10^{-4}$, so that any confidence drop relative to the reference model is counted as degradation. The constraint threshold $\varepsilon$ in Eq.~\ref{eq:constrained_obj} then controls the maximal tolerable average degradation over the training data. We set $\varepsilon = 1\times 10^{-4}$ by default, which effectively enforces that, on average, the knowledge-conditioned model does not perform worse than the reference model. We found that the LWGR is robust when $\delta$ and $\varepsilon$ are chosen within the range $[10^{-4}, 10^{-3}]$, and we report results with $\delta = \varepsilon = 1\times 10^{-4}$.

\subsection{Overall Performance (RQ1)}
To answer RQ1, we compare LWGR with representative baselines of different types on the Amazon datasets and a large-scale industrial dataset. The results in Table~\ref{tab:beauty_toys_industry} lead to the following observations.
\begin{itemize}[noitemsep, topsep=0pt, leftmargin=*]
    \item \textbf{LWGR consistently achieves the best performance across all datasets and metrics.} The relative gains over the strongest baseline reach up to 10.28\% in R@5 and 11.23\% in N@5. Across the three datasets, LWGR outperforms the best competing method by roughly 7\%–11\% across R@5/10 and N@5/10, and the improvement is positive for all metrics on all datasets.
    % Our method consistently achieves the best performance across all evaluation metrics, with the largest improvement reaching 8.58\%. These results indicate that semantic knowledge extracted from the LLM can effectively compensate for the limitations of conventional RS in modeling user intent, leading to both robust overall gains and a substantially higher performance upper bound. This demonstrates that fine-grained semantic understanding and personalized knowledge fusion are particularly beneficial in complex recommendation scenarios.
    \item \textbf{Prompt-based knowledge fusion consistently improves over the GR backbone, but the gains are relatively modest compared with LWGR.} On all datasets, prompt-based methods (TIGER+KAR and TIGER+SeRALM) are stronger than their GR backbones (TIGER), confirming the usefulness of external world knowledge. However, their improvements over the underlying GR models are relatively modest (mostly within 2--6\% on R@10), and sometimes inconsistent across metrics. In contrast, LWGR further improves over the strongest prompt-based baseline by up to 10.20\% in R@10 and 10.07\% in N@10, which we attribute to (i) end-to-end learned user-specific soft instructions that align LLM knowledge with individual behaviors, and (ii) Lagrangian-constrained fusion that selectively integrates beneficial knowledge while explicitly controlling degradation.
    % \item Prompt-based knowledge fusion methods outperform generative recommendation models without world knowledge, indicating that external knowledge can substantially improve the upper bound of generative recommendation. However, such knowledge is not always beneficial and may even hurt performance in some cases. In contrast, our method consistently surpasses all existing baselines, mainly owing to two key innovations: (1) the end-to-end learned user personalized soft instructions enables the LLM to dynamically generate semantic knowledge that closely aligns with each user’s behavioral patterns. (2) The knowledge fusion process is explicitly controlled via adaptive constraints, allowing the model to selectively and dynamically absorb world knowledge, which in turn leads to stable and reliable gains.
\end{itemize}

\begin{table}[t]
\centering
\caption{The impact of LLM size and training strategy on performance and efficiency, where R@5 represents model performance and QPS of training stage represents efficiency.}
\label{tab:llm_size_strategy_qps_hr5}
\label{tab:llm_size_strategy_qps_hr5}
\resizebox{0.9\linewidth}{!}{
\begin{tabular}{c c | c c | c c}
\toprule
\textbf{Scale} & \textbf{Strategy} & \textbf{QPS} & \textbf{Change} & \textbf{R@5} & \textbf{Improv.} \\
\midrule
0.6B & Frozen & 2.435 & -       & 0.2019 & -      \\
0.6B & LoRA   & 1.913 & -21.44\% & 0.2041 & +1.09\% \\
1.7B & Frozen & 1.954 & -19.75\% & 0.2087 & +3.37\% \\
1.7B & LoRA   & 1.421 & -41.64\% & 0.2095 & +3.76\% \\
4B   & Frozen & 1.695 & -30.39\% & 0.2145 & +6.24\% \\
8B   & Frozen & 1.038 & -57.37\% & 0.2154 & +6.69\% \\
\bottomrule
\end{tabular}}
\vspace{-10pt}
\end{table}

\subsection{Scaling the LLM Scale (RQ2)}
To study how the LLM scale interacts with deployment strategies, we conduct a systematic scaling experiment under six configurations: Qwen3-\{0.6B, 1.7B, 4B, 8B\} model sizes, combined with either frozen parameters or LoRA-based adaptation. We report both performance (R@5) and training throughput, measured by the number of training samples processed per second (QPS), on the industrial dataset in Table~\ref{tab:llm_size_strategy_qps_hr5}. We summarize three key observations:

\begin{itemize}[noitemsep, topsep=0pt, leftmargin=*]
  \item Larger LLMs bring consistent quality gains at the cost of lower training throughput. When keeping the LLM frozen, increasing the scale from 0.6B to 1.7B, 4B, and 8B monotonically improves R@5 , with the 8B–Frozen setting achieving up to 6.69\% relative improvement in R@5 over the 0.6B–Frozen baseline. Meanwhile, the training QPS decreases from 2.435 to 1.038 as the model scales from 0.6B to 8B, corresponding to a 57.37\% reduction at 8B compared with the 0.6B–Frozen setting. This confirms that, within LWGR, scaling up the LLM enhances recommendation quality by providing richer world knowledge and stronger semantic priors, but also introduces a clear efficiency–effectiveness trade-off.

  \item LoRA adaptation provides extra gains over frozen models of the same size. For both 0.6B and 1.7B LLMs, adding LoRA consistently improves R@5, but reduces QPS by 21.44\% and 41.64\% relative to frozen counterparts. This shows that a small number of task-specific parameters can better align generic LLM knowledge with recommendation signals, though the marginal accuracy gains must be weighed against the added training cost.

  \item Frozen large LLMs already provide a strong option under resource constraints. Across scales, the 4B–Frozen and 8B–Frozen settings outperform all 0.6B and 1.7B variants, including those with LoRA, while avoiding the complexity of adapting large models. This shows that LWGR can effectively exploit high-capacity frozen LLMs: simply increasing the frozen model size yields substantial gains over smaller tuned models, offering a practical route to better recommendation under limited training budgets.
\end{itemize}

\begin{table}[t]
\centering
\caption{Ablation study of LWGR on Industry.}
\label{tab:ablation_lwgr_industry}
\resizebox{0.85\linewidth}{!}{
\begin{tabular}{c| cc |cc}
\toprule
\textbf{Setting} & \textbf{R@5} &  \textbf{Improv.} & \textbf{N@5} & \textbf{Improv.} \\
\midrule
LWGR              & \textbf{0.2145} & -    & \textbf{0.1416} & - \\
LWGR$_{w/o\_Cons}$ & 0.2117 & -1.31\%   & 0.1399 & -1.20\% \\
LWGR$_{w/o\_Fus}$  & 0.2103 & -1.96\%   & 0.1375 & -2.90\% \\
LWGR$_{w/o\_PCB}$  & 0.2061 & -3.92\%   & 0.1346 & -4.94\% \\
LWGR$_{RQ}$        & 0.2120 & -1.17\% & 0.1402 & -0.99\% \\
LWGR$_{\beta}$ & 0.2118  & -1.26\% & 0.1391  & -1.77\% \\
\bottomrule
\end{tabular}}
\vspace{-14pt}
\end{table}

\subsection{Ablation Study (RQ3)}
In this section, we investigate the respective roles of five key components in LWGR on the industrial dataset:
\begin{itemize}[noitemsep, topsep=0pt, leftmargin=*]
    \item \textbf{LWGR$_{w/o\_PCB}$}: This variant removes the parallel codebooks for personalized soft instructions and instead feeds an MLP-projected user context concatenated with text embeddings into the LLM.
    \item \textbf{LWGR$_{RQ}$}: This variant replaces the parallel codebooks  with a residual quantization (RQ) scheme \cite{tiger,cobra}.
    \item \textbf{LWGR$_{w/o\_Fus}$}: This variant disables the cross-attention based knowledge fusion module and, as in the pilot experiments, directly concatenates the knowledge vectors to the GR encoder.
    \item \textbf{LWGR$_{w/o\_Cons}$}: This variant removes the inequality-constrained optimization with Lagrange multipliers and trains LWGR without any constraint on the performance gap to the reference model.
    \item \textbf{LWGR$_{\beta}$}: This variant replaces the adaptive Lagrangian optimization with a fixed penalty coefficient, using Eq.~(\ref{eq:manual_penalty}). We tune $\beta$ over $\{0.1, 0.3, 0.5, 0.7\}$  and report the best-performing setting.

\end{itemize}
From Table~\ref{tab:ablation_lwgr_industry}, we see that all key components contribute non-trivially to the final performance: removing any of them leads to clear drops in both R@5 and N@5. The largest degradation comes from \textbf{LWGR$_{w/o\_PCB}$}, where R@5 and N@5 decrease by 3.92\% and 4.94\%, respectively. This highlights that the codebook-based personalized soft instructions are crucial for capturing stable user intent and steering the LLM to produce preference-aligned knowledge; simply concatenating raw user context with text embeddings is far less effective.  Compared with LWGR, \textbf{LWGR$_{RQ}$} also lags behind, indicating that residual quantization is less effective than parallel codebook in modeling fine-grained user preferences. 

Ablating the fusion module in \textbf{LWGR$_{w/o\_Fus}$} also harms performance (R@5: \(-1.96\%\), N@5: \(-2.90\%\)), indicating that directly injecting LLM outputs as extra encoder features is inferior to our cross-attention-based fusion, where GR representations explicitly attend to world knowledge. Finally, \textbf{LWGR$_{w/o\_Cons}$} yields smaller but consistent drops, suggesting that the Lagrangian-based inequality constraint plays an important role in stabilizing ranking quality by preventing harmful knowledge injection. We further compare it with the fixed-penalty variant \textbf{LWGR$_{\beta}$}, whose R@5 and N@5 remain below LWGR, indicating that a fixed penalty achieves a weaker trade-off than the adaptive Lagrangian update. 
Overall, these ablations demonstrate that LWGR benefits from the joint effect of personalized soft instructions, structured cross-attention fusion, and constraint-guided optimization, and that removing any component noticeably undermines effectiveness.

\subsection{Analysis Experiments (RQ4)}
\begin{figure}[t]
    \vspace{-5pt}
    \captionsetup{skip=2pt} 
  \includegraphics[width=0.49\textwidth]{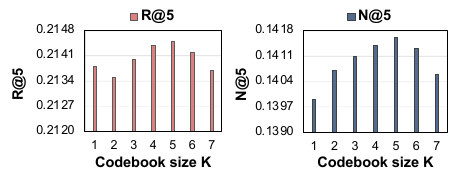}
  \caption{R/N@5 change under different codebook sizes $K$.}
  \label{fig:sensitive_k}
    \vspace{-14pt}
\end{figure}

In this section, we investigate how the number of sub-codebooks $K$ in the parallel codebook, which also corresponds to the number of personalized discrete codes per user, affects performance on the industrial dataset. The results are shown in Figure~\ref{fig:sensitive_k}. As $K$ increases from 1 to 5, both R@5 and N@5 steadily improve (by roughly 0.8\% and 1.2\%, respectively), indicating that a moderate increase in the number of personalized codes helps the model capture user behavior patterns with finer granularity. When $K$ is further increased beyond 5, both metrics start to drop slightly (by about 0.3\%–0.7\% relative to $K=5$), suggesting that too many sub-codebooks introduce redundancy and make the discrete representation harder to learn from sparse interactions. Based on this trade-off, we adopt $K=5$ as the default setting.
% In this section, we investigate how the number of sub-codebooks $K$ in the parallel codebook, which also corresponds to the number of personalized discrete codes assigned to each user, influences performance on the industrial dataset. The results are shown in Figure_4. As $K$ increases from 1 to 5, both HR@5 and NDCG@5 exhibit consistent improvements: HR@5 increases by approximately 0.8\%, and NDCG@5 improves by about 1.2\%. These results indicate that, up to a moderate level, allocating more personalized codes per user enables the model to capture user behavior patterns with finer granularity. When $K$ is further increased beyond 5, both metrics start to decline slightly, with a degradation of around 0.3\%--0.7\% compared with $K=5$. This suggests that an excessive number of sub-codebooks introduces redundancy and makes the discrete representation harder to learn from sparse interactions. Therefore, we adopt $K=5$ as the default setting.

\subsection{Online Experiments (RQ5)}
\label{sec:online_Experiments}
To further validate LWGR, we conduct an online A/B test on an advertising recommendation platform of a leading e-commerce company in Southeast Asia from Dec 19 to 30, 2025. The control group is based on TIGER \cite{tiger}, whereas the experimental group employs LWGR. Both groups each consist of 15\% randomly selected users. During the test window, we observe a \textbf{1.35\%} increase in \textbf{advertising revenue}, a \textbf{0.83\%} increase in \textbf{gross merchandise volume (GMV)}, and a \textbf{1.17\%} increase in \textbf{CTR} over the baseline. All these improvements pass a standard two-sided significance test with $p < 0.05$. With our proposed lightweight deployment scheme (Section~\ref{rec:online_deployment}), the average latency per request increased modestly from 13.35\,ms to 13.55\,ms (a relative increase of about 1.5\%) due to the additional vector retrieval and request forwarding. These online results further demonstrate that LWGR can bring tangible business gains when deployed in a large-scale industrial system.

\section{Conclusion}
In this paper, we propose LWGR, a framework that incorporates user personalized world knowledge into generative recommendation under Lagrangian constraints. We first systematically analyze the potential of LLM-generated semantic knowledge for improving recommendation performance, and identify two key challenges in generative recommendation: insufficient sensitivity to user interests and randomness in the performance gains. 
To address these challenges, LWGR consists of a knowledge extraction and a knowledge fusion component. The extraction module quantizes user context via a parallel codebook, converting user behaviors into personalized discrete codes that are fed into an LLM to obtain user personalized world knowledge. The fusion module then integrates this knowledge into the GR decoder through cross-attention. To prevent performance degradation, we cast fusion as an inequality-constrained optimization problem and apply a Lagrangian primal–dual method for adaptive integration.
Experiments on multiple datasets demonstrate that LWGR consistently and significantly outperforms strong baselines across all metrics, confirming the crucial role of personalized world knowledge in generative recommendation. In future work, we plan to investigate more efficient deployment strategies for larger-scale LLMs, as well as more interpretable and controllable collaboration mechanisms between LLMs and GR.

\bibliographystyle{ACM-Reference-Format}
\bibliography{reference}

\end{document}